\documentclass[prX,showpacs,nofootinbib]{revtex4}
\usepackage{amssymb}  

\newcommand{\be}{\begin{equation}}
\newcommand{\ee}{\end{equation}}
\newcommand{\ba}{\begin{eqnarray}}
\newcommand{\ea}{\end{eqnarray}}

\def\a{\alpha}
\def\b{\beta}
\def\d{\delta}
\def\e{\epsilon}
\def\ve{\varepsilon}
\def\f{\phi}
\def\vf{\varphi}

\def\h{\eta}

\def\l{\lambda}
\def\m{\mu}
\def\n{\nu}

\def\p{\pi}

\def\r{\rho}

\def\t{\tau}

\def\D{\Delta}
\def\F{\Phi}
\def\G{\Gamma}

\def\O{\Omega}
\def\P{\Pi}
\def\Q{\Theta}
\def\S{\Sigma}

\def\ca{{\cal A}}

\def\cg{{\cal G}}

\def\cj{{\cal J}}
\def\ck{{\cal K}}

\def\cp{{\cal P}}

\def\cs{{\cal S}}

\def\cw{{\cal W}}

\newcommand{\ov}{\overline}
\newcommand{\uv}{\underline}
\newcommand{\ti}{\tilde}
\newcommand{\wt}{\widetilde}

\newcommand{\aand}{\;\;\;\mbox{and}\;\;\;}
\newcommand{\pa}{\partial}

\def\bc{\bar c}
\def\bx{\bar\xi}
\newcommand{\la}{\langle}
\newcommand{\ra}{\rangle}

\def\I{\leavevmode\hbox{\small1\kern-3.8pt\normalsize1}}

\begin{document}

\title{The Jackiw-Pi model: classical theory}
\author{O.M. Del Cima}
\email{oswaldo.delcima@ufv.br}
\affiliation{Universidade Federal de Vi\c cosa (UFV),\\
Departamento de F\'\i sica - Campus Universit\'ario,\\
Avenida Peter Henry Rolfs s/n - 36570-000 -
Vi\c cosa - MG - Brazil.}
\date{\today}

\begin{abstract}
The massive even-parity non-Abelian gauge model in three 
space-time dimensions proposed by Jackiw and Pi is studied at the tree-level. The propagators are 
computed and the spectrum consistency is analyzed, besides, the symmetries of the 
model are collected and established through BRS invariance and Slavnov-Taylor identity. 
In the Landau gauge, thanks to the antighost equations and the Slavnov-Taylor identity, two rigid 
symmetries are identified by means of Ward identities. It is presented here a promising path for 
perturbatively quantization of the Jackiw-Pi model and a hint concerning its possible quantum scale invariance is 
also pointed out.
\\

\centerline{In honor of the 70th birthday of Prof. Olivier Piguet}
  
\end{abstract}
\maketitle

\section{Introduction}

The study of gauge field theories in three space-time dimensions has 
raised a great deal of interest since the early works of Deser, Jackiw and Templeton~\cite{deser-jackiw-templeton}. 
Over the last decades, this issue has also been motivated and well-supported 
in view of the possibilities they open up for the setting of a gauge field theoretical foundation
in the description of condensed matter phenomena, such as high-$T_{\rm c}$ superconductivity and 
quantum Hall effect. Meantime, one of the central problems in the framework of gauge field theories is the issue of gauge field mass. Gauge
symmetry is not, in principle, conflicting with the presence of a massive gauge boson. In two space-time
dimensions, the well-known Schwinger model puts in evidence the presence of a massive photon without the
breaking of gauge symmetry~\cite{schwinger}. Another evidence for the compatibility between gauge symmetry and massive
vector fields has been arisen in the study of three-dimensional gauge theories, when a topological mass term referred to as
the Chern-Simons one, once added to the Yang-Mills term, shifts the photon mass to a non-vanishing
value without breaking gauge invariance, however parity symmetry is lost~\cite{deser-jackiw-templeton}. 
Nevertheless, Jackiw and Pi overcame the challenge to implement both gauge and parity invariance in three space-time 
dimensions by breaking the Yang-Mills paradigm - non-Abelian generalizations of Abelian models. They proposed a  
three-dimensional non-Yang-Mills gauge model for a pair of vector fields with opposite parity 
transformations, which generates a mass-gap through a mixed Chern-Simons-like term preserving parity~\cite{jackiw-pi}. 
The Jackiw-Pi model has also been studied in the Hamiltonian framework~\cite{dayi}, where physical states consistency 
was demonstrated~\cite{dayi}. Recently, by using the BRS approach, new symmetries and 
gauge-fixing were established~\cite{delcima}, and in~\cite{gupta-kumar-malik} the Yang-Mills symmetry sector 
was analyzed through the Bonora-Tonin superfield formalism~\cite{bonora-tonin}. More recently, the authors of~\cite{deser-ertl-grumiller} 
have shown, by using the Hamiltonian formalism, that the bifurcation effect (a clash between two local invariances), stems for 
the three-dimensional Schouten-ghost-free gravity, there they also conjecture that such a bifurcation effect could 
appears in the Jackiw-Pi model -- since it presents two local invariances. 
The Jackiw-Pi model remains unquantized up to now, however, it is presented here the key ingredients for its further 
perturbatively quantization~\cite{jackiwpiquantum} through the algebraic method of renormalization~\cite{piguet}. 
In this work, the non-Abelian gauge model proposed by Jackiw and Pi, which generates
an even-parity mass term in three space-time dimensions, is revisited. The model and its gauge symmetries, the BRS symmetry, 
the gauge-fixing and the anti-fields action are presented in Section II. The BRS approach has allowed to 
bypass the difficulties addressed in the literature with respect to the gauge-fixing. In Section III, the 
tree-level propagators are computed, the spetrum consistency (causality and unitarity) is analyzed and the ultraviolet and infrared 
dimensions of all the fields are established. In the Section IV, the Slavnov-Taylor identity, ghost and 
anti-ghost equations, and the operatorial algebra are presented. Furthermore, it is shown that in the Landau gauge, 
thanks to the antighost equations and the Slavnov-Taylor identity, two rigid symmetries of the Jackiw-Pi model are 
identified by means of Ward identities.   

\section{The model and its symmetries}

\subsection{The model}

The classical action of the Jackiw-Pi model~\cite{jackiw-pi} is given by:
\be
\S_{\rm inv}={\rm Tr}\int{d^3 x} \left\{
{1\over2}F^{\m\n}F_{\m\n} 
+ {1\over2}\bigl(G^{\m\n}+g[F^{\m\n},\r]\bigr)\bigl(G_{\m\n}+g[F_{\m\n},\r]\bigr) 
- m\e^{\m\n\r}F_{\m\n} \f_\r \right\}~,
\label{action}
\ee
such that,
\be
F_{\m\n}=\pa_\m A_\n - \pa_\n  A_\m + g[A_\m,A_\n]~,~~G_{\m\n}=D_\m \f_\n - D_\n \f_\m 
\aand D_\m\bullet=\pa_\m\bullet + g[A_\m,\bullet~]~,
\ee
where $A_\m$ and $\f_\m$ are vector fields, $\r$ is a scalar, $g$ is a coupling constant and $m$ a mass parameter, also, $\bullet$ means any field. The Lie group is a simple compact, so that every field, $X=X_a\t_a$, is Lie algebra valued, with the matrices $\t$ being the generators of the group in the adjoint representation and obey 
\be
[\t_a,\t_b]=f_{abc}\t_c \aand {\rm Tr}(\t_a\t_b)=-{1\over2}\d_{ab}~~(a,b,c=1,2,\ldots,N^2-1)~.
\ee

\subsection{Gauge symmetries} 

The action (\ref{action}) is invariant under two sets of gauge transformations, $\d_\theta$ and $\d_\chi$:
\ba
&&\d_\theta A_\m=D_\m\theta ~,~~ \d_\theta \f_\m=g[\f_\m,\theta] \aand \d_\theta \r=g[\r,\theta] ~;\label{theta}\\
&&\d_\chi A_\m=0 ~,~~ \d_\chi \f_\m=D_\m\chi  \aand \d_\chi \r=-\chi ~,\label{chi}
\ea
where $\theta$ and $\chi$ are Lie algebra valued infinitesimal local parameters.

\subsection{BRS symmetry} 

The corresponding BRS transformations of the fields $A_\m$, $\f_\m$ and $\r$, stemming from the symmetries 
(\ref{theta}) and (\ref{chi}), are given by\footnote{The commutators among the fields are assumed to 
be graded, namely, $[\vf_1^{g_1},\vf_2^{g_2}]
\equiv\vf_1^{g_1}\vf_2^{g_2}-(-1)^{g_1.g_2}\vf_2^{g_2}\vf_1^{g_1}$, where the 
upper indices, $g_1$ and $g_2$, are the Faddeev-Popov charges ($\F\P$) carried 
by the fields $\vf_1^{g_1}$ and $\vf_2^{g_2}$, respectively.}:
\ba
&& sA_\m=D_\m c ~,~~ s\f_\m=D_\m \xi + g[\f_\m,c] ~,~~ s\r=-\xi + g[\r,c]~, \nonumber\\
&& sc=-gc^2 \aand s\xi=-g[\xi,c]~, \label{BRS}
\ea
where $c$ and $\xi$ are the Faddeev-Popov ghosts, with Faddeev-Popov charge 
(ghost number) one. The ghost number ($\F\P$) of all 
fields and antifields are collected in Table~\ref{dimensions}

\subsection{The gauge-fixing and the antifields action} 

The gauge-fixing adopted here belongs to the class of the linear covariant gauges discussed 
by 't Hooft~\cite{thooft}. In order to implement the gauge-fixing following the BRS procedure~\cite{brs}, we introduce two sorts of ghosts ($c$ and $\xi$), antighosts ($\bc$ and $\bx$) and the Lautrup-Nakanishi fields~\cite{lautrup-nakanishi} ($b$ and $\p$), playing the role of Lagrange multiplier fields 
for the gauge condition, such that  
\ba
&&s\bc=b~,~~sb=0~;  \label{doubletbc}\\
&&s\bx=\p~,~~s\p=0~;\label{doubletbd}
\ea
where the multiplier fields, $b$ and $\p$, and the Faddeev-Popov antighosts, $\bc$ and $\bx$, with ghost number minus one, belong to the BRS-doublets (\ref{doubletbc}) and (\ref{doubletbd}).

Now, by adopting the gauge conditions
\ba
\frac{\d \S_{\rm{gf}}}{\d b}= \pa^\m A_\m + \a b~,\label{gaugefixing1}\\
\frac{\d \S_{\rm{gf}}}{\d \p}= \pa^\m \f_\m + \b \p~,\label{gaugefixing2}
\ea
it follows that the BRS-trivial gauge-fixing action compatible with then reads
\ba
\S_{\rm{gf}}&\!\!=\!\!&s~{\mbox{Tr}}\int d^3x~\left\{ \bc\pa^\m A_\m + \bx
\pa^\m \f_\m\ + \frac{\a}{2} \bc b + \frac{\b}{2}\bx \p \right\} \nonumber\\
&\!\!=\!\!&{\mbox{Tr}}\int d^3x~\left\{ b\pa^\m A_\m - \bc \pa^\m D_\m c 
+ \p \pa^\m \f_\m\ - \bx \pa^\m \bigl( D_\m \xi + g[\f_\m,c] \bigr) 
+ \frac{\a}{2} b^2 + \frac{\b}{2} \p^2 \right\}~.
\ea
Let us now introduce the action in which the nonlinear BRS transformations are coupled to the antifields 
(BRS invariant external fields), so as to control, at the quantum level, the renormalization of those transformations:
\be
\S_{\rm{ext}}={\mbox{Tr}}\int d^3x~\left\{ A^*_\m sA^\m + \f^*_\m s\f^\m 
+ \r^* s\r + c^* sc + \xi^* s\xi \right\}~, \label{sigmaext}
\ee
where, as mentioned above, the antifields are BRS invariant, namely, 
\be
sA^*_\m=s\f^*_\m=s\r^*=sc^*=s\xi^*=0~.\label{BRSanti}
\ee

The total action at the tree level for the Jackiw-Pi model, $\G^{(0)}$, 
is therefore given by:
\be
\G^{(0)}=\S_{\rm inv} + \S_{\rm{gf}} + \S_{\rm{ext}}~,\label{totalaction}
\ee
which is invariant under the BRS transformations given by the equations (\ref{BRS}), (\ref{doubletbc}), (\ref{doubletbd}) and (\ref{BRSanti}). The action (\ref{totalaction}) preserves the ghost number. The values of the ghost number, the ultraviolet (UV) and the infrared (IR) dimensions (respected 
to the Landau gauge) are displayed in Table~\ref{dimensions} - all subtleties concerning the determination of the UV and the IR dimensions of the fields, in the Landau gauge, is presented in the next section. The statistics is defined as follows: the fields 
of integer spin and odd ghost number as well as the fields of half integer spin and even ghost number are 
anticommuting; the other fields commute with the formers and among themselves.

An interesting feature of the Jackiw-Pi action $\G^{(0)}$(\ref{totalaction}) is that it is not BRS local 
invariant thanks to the parity-even mass term:
\be
\S_{\rm m}={\rm Tr}\int{d^3 x} \left\{- m\e^{\m\n\r}F_{\m\n} \f_\r \right\}~,
\ee 
since
\be
sF_{\m\n}=g[F_{\m\n},c]~,
\ee
then
\be
s\S_{\rm m}=-m~s{\rm Tr}\int{d^3 x} \left\{\e^{\m\n\r}F_{\m\n} \f_\r \right\}= 
-m~{\rm Tr}\int{d^3 x} \left\{ \e^{\r\m\n}\pa_\r(F_{\m\n}\xi) \right\}~,
\ee
which is invariant only up to a total derivative, possibly indicating that at the quantum level the $\b$-function 
associated to the mass parameter $m$ vanishes~\cite{delcima-franco-helayel-piguet, barnich}.

\section{Spectral analysis}

In quantum field theory, unitarity and causality are essential physical requirements. Unitarity (of the $S$-matrix) 
reflects the fundamental principle of probability conservation -- meaning the absence of  
negative-norm 1-particle states in the spectrum. Even though we have to introduce in certain instances the artificial 
device of an indefinite metric in Hilbert space, the physical quantities always refer to positive-norm states, preserved 
through the time evolution. Causality principle establishes a time correlation among the cause and its subsequent effect, 
requiring that the change in the interaction law in any space-time region can influence the evolution of the system only at 
subsequent times.

\subsection{The propagators} 

The propagators are the key ingredient to the analysis of the spectral consistency and the unitarity at 
the tree-level of the model, as well as  
in the determination of the ultraviolet ($d$) and infrared ($r$) dimensions of the fields.

By switching off the coupling constant $g$ we get the free part of the action, $\S_{\rm inv} + \S_{\rm{gf}}$, as 
follows:
\ba
\S_{\rm free}&=&{\rm Tr}\int{d^3 x}~\biggl\{{1\over2}F^{\m\n}F_{\m\n} + {1\over2}G^{\m\n}G_{\m\n} 
- m\e^{\m\n\r}F_{\m\n} \f_\r + b\pa^\m A_\m + \frac{\a}{2} b^2 + \p \pa^\m \f_\m\ + \frac{\b}{2} \p^2 + \nonumber\\
&-& \bc \pa^\m \pa_\m c  - \bx \pa^\m \pa_\m \xi \biggr\}~,
\label{freeaction}
\ea
where, by means of the operators: 
\be
\Q^{\m\n}=\h^{\m\n}-\frac{\pa^\m\pa^\n}{\square}~,~~\O^{\m\n}=\frac{\pa^\m\pa^\n}{\square} 
\aand \S^{\m\n}=\e^{\m\r\n}\pa_\r~,
\ee
that fulfil the algebra displayed in Table~\ref{algebra}, the free action $\S_{\rm free}$(\ref{freeaction}) 
can be written as:
\ba
\S_{\rm free}&=&{\rm Tr}\int{d^3 x}~\biggl\{- A_\m\square\Q^{\m\n}A_\n - \f_\m\square\Q^{\m\n}\f_\n 
- 2m A_\m \S^{\m\n}\f_\n 
+ b\pa^\m A_\m + \frac{\a}{2} b^2 + \p \pa^\m \f_\m\ + \frac{\b}{2} \p^2 + \nonumber \\ 
&-& \bc\square c  - \bx\square \xi \biggr\}~,\nonumber\\
&=&\int{d^3 x}~\biggl\{{1\over2}A^a_\m\square\Q^{\m\n}A^a_\n + {1\over2}\f^a_\m\square\Q^{\m\n}\f^a_\n 
+ m A^a_\m \S^{\m\n}\f^a_\n 
- {1\over2}b^a\pa^\m A^a_\m - \frac{\a}{4} b^a b^a - {1\over2}\p^a \pa^\m \f^a_\m\ - \frac{\b}{4} \p^a\p^a + \nonumber\\ 
&+& {1\over2}\bc^a\square c^a + {1\over2}\bx^a\square \xi^a \biggr\}~.
\label{freeactionops}
\ea

The generating functional for the connected Green functions ($Z^{\rm c}[J]$) is defined by means 
of the vertex functional ($\G^{(0)}$) through the Legendre transformation~\cite{itzykson-zuber}:
\be
Z^{\rm c}[J_i] = \G^{(0)}[\F_i] + {\rm Tr}\int{d^3 x}~\left( A_\m J_A^\m + \f_\m J_\f^\m + bJ_b + \p J_\p  
+ \bar{J_c}c + J_{\bc}\bc + \bar{J_\xi}\xi + J_{\bx}\bx \right)~, 
\label{zc}
\ee
where $\F_i=(A_\m,\f_\m,b,\p,c,\bc,\xi,\bx)$ and $J_i=(J_A^\m,J_\f^\m,J_b,J_\p,\bar{J_c},J_{\bc},\bar{J_\xi},J_{\bx})$, 
such that
\ba
&&\frac{\d Z^{\rm c}}{\d J_A^\m(x)}=A_\m(x)~,~~\frac{\d \G^{(0)}}{\d A_\m(x)}=-J_A^\m(x)~,~~ 
\frac{\d Z^{\rm c}}{\d J_\f^\m(x)}=\f_\m(x)~,~~\frac{\d \G^{(0)}}{\d \f_\m(x)}=-J_\f^\m(x)~,\nonumber\\
&&\frac{\d Z^{\rm c}}{\d J_b(x)}=b(x)~,~~\frac{\d \G^{(0)}}{\d b(x)}=-J_b(x)~,~~ 
\frac{\d Z^{\rm c}}{\d J_\p(x)}=\p(x)~,~~\frac{\d \G^{(0)}}{\d \p(x)}=-J_\p(x)~,\nonumber\\
&&\frac{\d Z^{\rm c}}{\d \bar{J_c}(x)}=c(x)~,~~\frac{\d \G^{(0)}}{\d c(x)}=\bar{J_c}(x)~,~~ 
\frac{\d Z^{\rm c}}{\d J_{\bc}(x)}=\bc(x)~,~~\frac{\d \G^{(0)}}{\d \bc(x)}=J_{\bc}(x)~,\nonumber\\
&&\frac{\d Z^{\rm c}}{\d \bar{J_\xi}(x)}=\xi(x)~,~~\frac{\d \G^{(0)}}{\d \xi(x)}=\bar{J_\xi}(x)~,~~ 
\frac{\d Z^{\rm c}}{\d J_{\bx}(x)}=\bx(x)~,~~\frac{\d \G^{(0)}}{\d \bx(x)}=J_{\bx}(x)~. \label{zc-vertex}
\ea 

The tree-level propagators for all the fields:
\be
\la T \F_i(x) \F_j(y) \ra = - i \frac{\d^2 Z^{\rm c}}{\d J_i(x)\d J_j(y)}~,
\ee
are then computed, through the use of Eq.(\ref{zc-vertex}), as follows:
\ba
&& \la T A^a_\m(x) A^b_\n(y) \ra = - i \frac{\d A^a_\m(x)}{\d J_A^{b\n}(y)}~,~~
\la T \f^a_\m(x) \f^b_\n(y) \ra = - i \frac{\d \f^a_\m(x)}{\d J_\f^{b\n}(y)}~,~~
\la T A^a_\m(x) \f^b_\n(y) \ra = - i \frac{\d A^a_\m(x)}{\d J_\f^{b\n}(y)}~, \nonumber\\
&& \la T A^a_\m(x) b^b(y) \ra = - i \frac{\d A^a_\m(x)}{\d J_b^{b}(y)}~,~~
\la T \f^a_\m(x) \p^b(y) \ra = - i \frac{\d \f^a_\m(x)}{\d J_\p^{b}(y)}~, \nonumber\\ 
&& \la T b^a(x) b^b(y) \ra = - i \frac{\d b^a(x)}{\d J_b^{b}(y)}~,~~
\la T \p^a(x) \p^b(y) \ra = - i \frac{\d \p^a(x)}{\d J_\p^{b}(y)}~, \nonumber\\
&& \la T c^a(x) \bc^b(y) \ra = i \frac{\d c^a(x)}{\d J_{\bc}^{b}(y)}~,~~
\la T \xi^a(x) \bx^b(y) \ra = i \frac{\d \xi^a(x)}{\d J_{\bx}^{b}(y)}~. \label{propeqs}
\ea
It should be noticed that the functional derivatives satisfy the following property:
\be
\frac{\d^2}{\d X_1^{g_1}(x)\d X_2^{g_2}(y)}=(-1)^{g_1.g_2}\frac{\d^2}{\d X_2^{g_2}(y)\d X_1^{g_1}(x)}~,
\ee
where the upper indices, $g_1$ and $g_2$, are the Faddeev-Popov charges ($\F\P$) carried by the fields or 
currents, $X_1^{g_1}$ and $X_2^{g_2}$, respectively. Due to the fact that the functional $Z^{\rm c}[J]$ (\ref{zc}) has ghost 
number zero, the ``classical'' sources $J_i=(J_A^\m,J_\f^\m,J_b,J_\p,\bar{J_c},J_{\bc},\bar{J_\xi},J_{\bx})$ into the Legendre transformation (\ref{zc}), which relates the 
connected functional $Z^{\rm c}[J]$ and vertex functional $\G^{(0)}[\F]$, have ghost numbers 
$\F\P(J_i)=(0,0,0,0,-1,1,-1,1)$. 

\begin{table}[t]
\begin{center}
\begin{tabular}{|c|c|c|c|}
\hline
& $\Theta_{\l\n}$ & $\Omega_{\l\n}$ & $\S_{\l\n}$ \\ \hline
$\Theta^{\m\l}$ & $\Theta^\m_{~\n}$ & $0$ & $\S^\m_{~\n}$ \\ \hline
$\Omega^{\m\l}$ & $0$ & $\Omega^\m_{~\n}$ & $0$ \\ \hline
$\S^{\m\l}$ & $\S^\m_{~\n}$ & $0$ & $-\square\Theta^\m_{~\n}$ \\ \hline
\end{tabular}
\end{center}
\caption[]{Operator algebra fulfilled by $\Theta$, $\Omega$ and $\S$\label{algebra}.}
\end{table}

From the equations of motion we get, $J_i\equiv J_i[\F_j]$:
\ba
&&\frac{\d \G^{(0)}}{\d A^a_\m} = \square\Q^{\m\n}A^a_\n + m\S^{\m\n}\f^a_\n + {1\over2}\pa^\m b^a = -J_A^{a\m}~,~~
\frac{\d \G^{(0)}}{\d b^a}= - {1\over2}\pa^\m A^a_\m - \frac{\a}{2} b^a = -J_b^a~, \nonumber\\
&&\frac{\d \G^{(0)}}{\d \f^a_\m} = \square\Q^{\m\n}\f^a_\n + m\S^{\m\n}A^a_\n + {1\over2}\pa^\m \p^a = -J_\f^{a\m}~,~~
\frac{\d \G^{(0)}}{\d \p^a}= - {1\over2}\pa^\m \f^a_\m - \frac{\b}{2} \p^a = -J_\p^a~, \nonumber\\
&&\frac{\d \G^{(0)}}{\d \bc^a}= {1\over2}\square c^a = J_{\bc}^{a}~,~~
\frac{\d \G^{(0)}}{\d \bx^a}= {1\over2}\square\xi^a = J_{\bx}^{a}~, \label{eqmotion}
\ea
where by solving these equations of motion (\ref{eqmotion}) so as to express, $\F_i\equiv \F_i[J_j]$, and 
adopting the algebra fulfilled by the operators $\Q_{\m\n}$, $\O_{\m\n}$ and $\S_{\m\n}$ displayed 
in Table~\ref{algebra}, it is found that:
\ba
&&A^a_\m = - \bigg\{\frac{1}{\square + m^2}\Q_{\m\n} - \frac{2\a}{\square}\O_{\m\n}\bigg\}J_A^{a\n} + \frac{m}{\square(\square + m^2)}\S_{\m\n}J_\f^{a\n} + \frac{2}{\square} \pa_\m J_b^{a}~, \nonumber \\
&&\f^a_\m = - \bigg\{\frac{1}{\square + m^2}\Q_{\m\n} - \frac{2\b}{\square}\O_{\m\n}\bigg\}J_\f^{a\n} + \frac{m}{\square(\square + m^2)}\S_{\m\n}J_A^{a\n} + \frac{2}{\square} \pa_\m J_\p^{a}~, \nonumber \\ 
&& b^a = -\frac{2}{\square}\pa_\m J_A^{a\m}~,~~ \p^a = -\frac{2}{\square}\pa_\m J_\f^{a\m}~,~~
c^a = \frac{2}{\square}J_{\bc}^{a}~,~~ \xi^a = \frac{2}{\square}J_{\bx}^{a}~. \label{fields}
\ea

Now, by substituting the fields solutions, presented above in Eq.(\ref{fields}), into those ones in Eq.(\ref{propeqs}), 
the tree-level propagators are given by: 
\ba
&& \la T A^a_\m(x) A^b_\n(y) \ra = i \d^{ab} \bigg\{\frac{1}{\square + m^2}\Q_{\m\n} - \frac{2\a}{\square}\O_{\m\n}\bigg\}\d^3(x-y)~, \nonumber\\
&&\la T \f^a_\m(x) \f^b_\n(y) \ra = i \d^{ab} \bigg\{\frac{1}{\square + m^2}\Q_{\m\n} - \frac{2\b}{\square}\O_{\m\n}\bigg\}\d^3(x-y)~, \nonumber\\
&&\la T A^a_\m(x) \f^b_\n(y) \ra = - i \d^{ab} \frac{m}{\square(\square + m^2)}\S_{\m\n}\d^3(x-y)~, \nonumber\\
&& \la T A^a_\m(x) b^b(y) \ra = - i \d^{ab} \frac{2}{\square}\pa_\m\d^3(x-y)~,~~
\la T \f^a_\m(x) \p^b(y) \ra = - i \d^{ab} \frac{2}{\square}\pa_\m\d^3(x-y)~, \nonumber\\ 
&& \la T b^a(x) b^b(y) \ra = 0~,~~\la T \p^a(x) \p^b(y) \ra = 0~, \nonumber\\
&& \la T c^a(x) \bc^b(y) \ra = i \d^{ab} \frac{2}{\square}\d^3(x-y)~,~~
\la T \xi^a(x) \bx^b(y) \ra = i \d^{ab} \frac{2}{\square}\d^3(x-y)~, \label{prop}
\ea
where, assuming
\be
\d^3(x-y)=\int \frac{d^3k}{(2\p)^3}~e^{-ik(x-y)}~,
\ee
the propagators in momenta space read:
\ba
&& \la A^a_\m(k) A^b_\n(k) \ra = -i \d^{ab} \bigg\{\frac{1}{k^2 - m^2}\left(\h_{\m\n}-\frac{k_\m k_\n}{k^2}\right) - \frac{2\a}{k^2}\left(\frac{k_\m k_\n}{k^2}\right)\bigg\}~, \label{propkAA}\\
&&\la \f^a_\m(k) \f^b_\n(k) \ra = -i \d^{ab} \bigg\{\frac{1}{k^2 - m^2}\left(\h_{\m\n}-\frac{k_\m k_\n}{k^2}\right) - \frac{2\b}{k^2}\left(\frac{k_\m k_\n}{k^2}\right)\bigg\}~, \label{propkff}\\
&&\la A^a_\m(k) \f^b_\n(k) \ra = - \d^{ab} \frac{m}{k^2(k^2 - m^2)}\e_{\m\r\n}k^\r~, \label{propkAf}\\
&& \la A^a_\m(k) b^b(k) \ra = \d^{ab} \frac{2}{k^2}k_\m~,~~
\la \f^a_\m(k) \p^b(k) \ra = \d^{ab} \frac{2}{k^2}k_\m~, \label{propkAb}\\ 
&& \la b^a(k) b^b(k) \ra = 0~,~~\la \p^a(k) \p^b(k) \ra = 0~, \label{propkbb}\\
&& \la c^a(k) \bc^b(k) \ra = - i \d^{ab} \frac{2}{k^2}~,~~
\la \xi^a(k) \bx^b(k) \ra = - i \d^{ab} \frac{2}{k^2}~. \label{propkcbc}
\ea

\subsection{Unitarity and causality}

We will now discuss the spectrum and tree-level unitarity of the model. By coupling the propagators to external currents, $\cj_{\F_i}^a=(\cj^{a\m}_A,\cj^{a\m}_{\f},\cj^a_b,\cj^a_\p,\cj^a_c,\cj^a_{\bc},\cj^a_\xi,\cj^a_{\bx})$, compatible with the symmetries of the model, and then taking the imaginary part of the residues of the current-current 
amplitudes, $\ca_{\F_i\F_j}$, at the poles, we can probe the necessary conditions for unitarity (positive imaginary part of the residues of 
the transition amplitudes, $\Im{\rm Res}~\ca_{\F_i\F_j}>0$, as a consequence of the $S$-matrix be unitary) at the tree-level and count the degrees of freedom described by the fields, $\F^a_i=(A^a_\m,\f^a_\m,b^a,\p^a,c^a,\bc^a,\xi^a,\bx^a)$. 
The current-current transition amplitudes in momentum space are written as:
\be
\ca_{\F_i\F_j} = \cj_{\F_i}^{*a}(k) \la \F_i^a(k) \F_j^b(k) \ra \cj_{\F_j}^{b}(k)~.
\ee

At this moment we will first analyze the case of the propagators of the vector fields 
$A^a_\m$ and $\f^a_\m$, given by Eqs.($\ref{propkAA}$)--($\ref{propkAf}$). The vector currents, $\cj^{a\m}_A$ and $\cj^{a\m}_{\f}$, can be expanded in terms of a three-dimensional complete basis in the  momentum space as follows:
\be
\cj^{a\m}_A = X^a_A k^\m + Y^a_A {\ti k}^\m + Z^a_A \ve^\m \aand 
\cj^{a\m}_\f = X^a_\f k^\m + Y^a_\f {\ti k}^\m + Z^a_\f \ve^\m~,
\ee
fulfilling the current conservation conditions:
\be
k_\m\cj^{a\m}_A = 0 \aand k_\m\cj^{a\m}_\f = 0~, \label{kJ}
\ee
where $k^\m=(k^0,{\vec k})$, ${\ti k}^\m=(k^0,-{\vec k})$ and $\ve^\m=(0,{\vec \ve})$ are linearly independent 
vectors satisfying the constraints:
\be
k^\m\ve_\m = {\ti k}^\m\ve_\m = 0 \aand \ve^\m\ve_\m = -1~,
\ee
such that for a massive pole, $k^\m k_\m={\ti k}^\m{\ti k}_\m=m^2$, and for a massless one, 
$k^\m k_\m={\ti k}^\m{\ti k}_\m=0$. 

In the massive case ($k^2=m^2$), the momentum can be chosen as $k^\m=(m,{\vec 0})$, and by the current conservation conditions (\ref{kJ}), the currents $\cj^{a\m}_A$ and $\cj^{a\m}_\f$ are given by:
\be
\cj^{a\m}_A|_{k^2=m^2} = Z^a_A(0,{\vec \ve}) \aand \cj^{a\m}_\f|_{k^2=m^2} = Z^a_\f(0,{\vec \ve})~.\label{Jm} 
\ee 
On the other hand, in the massless case ($k^2=0$), the momentum chosen as $k^\m=(m,0,m)$ together with 
the current conservation conditions (\ref{kJ}) fix the currents $\cj^{a\m}_A$ and $\cj^{a\m}_\f$ as below: 
\be
\cj^{a\m}_A|_{k^2=0} = (mX^a_A,Z^a_A,mX^a_A) \aand \cj^{a\m}_\f|_{k^2=0} = (mX^a_\f,Z^a_\f,mX^a_\f)~.\label{J0}  
\ee 

The current-current amplitudes for the vector fields $A^a_\m$ and $\f^a_\m$ are given by:
\ba
&&\ca_{AA} = \cj^{*a\m}_A(k) \la A^a_\m(k) A^b_\n(k) \ra \cj^{b\n}_A(k) = 
-i\frac{1}{k^2 - m^2}~\cj^{*a\m}_A\cj^{a}_{A\m}~,\label{Aaa}\\
&&\ca_{\f\f} = \cj^{*a\m}_\f(k) \la \f^a_\m(k) \f^b_\n(k) \ra \cj^{b\n}_\f(k) = 
-i\frac{1}{k^2 - m^2}~\cj^{*a\m}_\f\cj^{a}_{\f\m}~,\label{Aff}\\
&&\ca_{A\f} = \cj^{*a\m}_A(k) \la A^a_\m(k) \f^b_\n(k) \ra \cj^{b\n}_\f(k) = 
-\frac{m}{k^2(k^2 - m^2)}~\e_{\m\r\n}\cj^{*a\m}_Ak^\r\cj^{a\n}_\f~,\label{Aaf}
\ea
where use has been made of the current conservation conditions (\ref{kJ}). Analyzing the amplitudes above, 
it can be verified that the amplitudes $\ca_{AA}$ (\ref{Aaa}) and $\ca_{\f\f}$ (\ref{Aff}) have single massive 
poles at $k^2=m^2$, whereas the amplitude $\ca_{A\f}$ (\ref{Aaf}) has two poles, a massive and a massless, at 
$k^2=m^2$ and $k^2=0$, respectively. Bearing in mind the currents $\cj^{a\m}_A$ and $\cj^{a\m}_\f$ calculated 
at the poles $k^2=m^2$ (\ref{Jm}) and $k^2=0$ (\ref{J0}), the residues of the current-current amplitudes $\ca_{AA}$, $\ca_{\f\f}$ and $\ca_{A\f}$ evaluated at their respective poles give rise to:
\ba
&&{\rm Res}~\ca_{AA}|_{k^2=m^2} = i |Z^a_A|^2~,\\
&&{\rm Res}~\ca_{\f\f}|_{k^2=m^2} = i |Z^a_\f|^2~,\\
&&{\rm Res}~\ca_{A\f}|_{k^2=m^2} = 0 \aand {\rm Res}~\ca_{A\f}|_{k^2=0} = 0~.
\ea 
Therefore, by considering the imaginary part of the residues ($\Im{\rm Res})$ at the poles, we get:
\ba
&&\Im{\rm Res}~\ca_{AA}|_{k^2=m^2} = |Z^a_A|^2>0~,\label{ImResAaa}\\
&&\Im{\rm Res}~\ca_{\f\f}|_{k^2=m^2} = |Z^a_\f|^2>0~,\label{ImResAff}\\
&&\Im{\rm Res}~\ca_{A\f}|_{k^2=m^2} = 0 \aand \Im{\rm Res}~\ca_{A\f}|_{k^2=0} = 0~,\label{ImResAaf}
\ea
then, it can be concluded from ($\ref{ImResAaa}$) and ($\ref{ImResAff}$) that the both vector fields, $A^a_\m$ and $\f^a_\m$, carry $2(N^2-1)$ massive degrees of freedom with mass $m$, however, from ($\ref{ImResAaf}$) it follows that there are no massless degrees of freedom propagating associated to the vector fields.

Let us now analize the propagators related to the fields $b^a$, $\p^a$, $c^a$, $\bc^a$, $\xi^a$ and $\bx^a$, 
given by Eqs.(\ref{propkAb})--(\ref{propkcbc}). The current-current amplitudes read:
\ba
&&\ca_{Ab} = \cj^{*a\m}_A(k) \la A^a_\m(k) b^b(k) \ra \cj^b_b(k) = 
\frac{2}{k^2}~k_\m\cj^{*a\m}_A\cj^a_b=0~,\label{Aab}\\
&&\ca_{\f\p} = \cj^{*a\m}_\f(k) \la \f^a_\m(k) \p^b(k) \ra \cj^b_\p(k) = 
\frac{2}{k^2}~k_\m\cj^{*a\m}_\f\cj^a_\p=0~,\label{Afp}\\
&&\ca_{bb} = \cj^{*a}_b(k) \la b^a(k) b^b(k) \ra \cj^b_b(k) = 0~,~~ 
\ca_{\p\p} = \cj^{*a}_\p(k) \la \p^a(k) \p^b(k) \ra \cj^b_\p(k) = 0~,\label{Abb}\\
&&\ca_{c \bc} = \cj^{*a}_{c}(k) \la c^a(k) \bc^b(k) \ra \cj^b_{\bc}(k) = - i \frac{2}{k^2}\cj^{*a}_{c}\cj^a_{\bc}~,~~ 
\ca_{\xi\bx} = \cj^{*a}_{\xi}(k) \la \xi^a(k) \bx^b(k) \ra \cj^b_{\bx}(k) = - i \frac{2}{k^2}\cj^{*a}_{\xi}\cj^a_{\bx}~,\label{Acbc}
\ea
where the current conservation conditions (\ref{kJ}) were applied in (\ref{Aab}) and (\ref{Afp}). Through the amplitudes displayed 
above, by considering their imaginary parts of the residues at the massless pole $k^2=0$:
\ba
&&\Im{\rm Res}~\ca_{Ab}|_{k^2=0} = 0~,~~ \Im{\rm Res}~\ca_{\f\p}|_{k^2=0} = 0~,
~~ \Im{\rm Res}~\ca_{bb}|_{k^2=0} = 0~,~~ \Im{\rm Res}~\ca_{\p\p}|_{k^2=0} = 0~,\label{ImResAAb}\\
&&\Im{\rm Res}~\ca_{c \bc}|_{k^2=0} = -2~\cj^{*a}_{c}\cj^a_{\bc}<0 \aand 
\Im{\rm Res}~\ca_{\xi\bx}|_{k^2=0} = -2~\cj^{*a}_{\xi}\cj^a_{\bx}<0~,\label{ImResAcc}
\ea
it shows that there are no massless modes propagating in the Lautrup-Nakanishi fields sector (\ref{ImResAAb}), nevertheless, 
from (\ref{ImResAcc}) we see that the massless propagating (negative norm state) Faddeev-Popov ghosts (antighots) $c^a$ and $\xi^a$ ($\bc^a$ and $\bx^a$) carry, each of them, $N^2-1$ degrees of freedom -- taking care of the $N^2-1$ spurious degrees of freedom stemming
from the longitudinal sector of each vector field, $A^a_\m$ (\ref{propkAA}) and $\f^a_\m$ (\ref{propkff}). 

From the results presented above, it can be concluded that the Jackiw-Pi model is free from tachyons and ghosts 
at the classical level. Nevertheless, to have full control of the unitarity at tree-level, it is still necessary to 
study the behaviour of the scattering cross sections in the limit of high center of mass energies, by analizing 
the Froissart-Martin bound~\cite{chaichian,delcima1}.

\subsection{Ultraviolet and infrared dimensions} 

In order to establish the 
ultraviolet (UV) and infrared (IR) dimensions of any field, 
$X$ and $Y$, we make use of the 
UV and IR asymptotical behaviour of their propagator, 
$\D_{XY}(k)$, $d_{XY}$ and $r_{XY}$, respectively: 
\be
d_{XY}={\ov{\rm deg}}_{k}\D_{XY}(k) \aand r_{XY}={\uv{\rm deg}}_{k}\D_{XY}(k)~,
\ee
where the upper degree ${\ov{\rm deg}}_{k}$ gives the asymptotic power 
for $k\rightarrow \infty$ whereas the lower degree ${\uv{\rm deg}}_{k}$ 
gives the asymptotic power for $k\rightarrow 0$. The UV ($d$) and IR ($r$) 
dimensions of the fields, $X$ and $Y$, are chosen to fulfill the 
following inequalities:
\be
d_X + d_Y \geq 3 + d_{XY} \aand r_X + r_Y \leq 3 + r_{XY}~. \label{uv-ir}
\ee

Since the Landau gauge shall be adopted later, the UV and IR dimensions of all the fields 
are fixed assuming $\a=\b=0$. In order to fix the UV and IR dimensions of the vector fields $A_\m$ and $\f_\m$, 
use has been made of the propagators, (\ref{propkAA}), (\ref{propkff}) and (\ref{propkAf}), together with the conditions 
(\ref{uv-ir}), and the following conditions stem from:
\ba
&& 2d_A\geq 1~,~~ 2d_\f\geq 1 \aand d_A + d_\f \geq 0 ~\longrightarrow~ d_A=d_\f=\frac{1}{2}~; \label{dAf}\\
&& 2r_A\leq 3~,~~ 2r_\f\leq 3 \aand r_A + r_\f \leq 2 ~\longrightarrow~ r_A=r_\f=\frac{1}{2}~. \label{rAf}
\ea
From the propagators (\ref{propkAb}) and the conditions, (\ref{uv-ir}), (\ref{dAf}) and (\ref{rAf}), we can fix the UV and IR 
dimensions of the Lautrup-Nakanishi fields, $b$ and $\p$, as follows:
\ba
&& d_A + d_b \geq 2 \aand d_A=\frac{1}{2} ~\longrightarrow~ d_b=\frac{3}{2}~;
~~d_\f + d_\p \geq 2 \aand d_\f=\frac{1}{2} ~\longrightarrow~ d_\p=\frac{3}{2}~; \\
&& r_A + r_b \leq 2 \aand r_A=\frac{1}{2} ~\longrightarrow~ r_b=\frac{3}{2}~;
~~r_\f + r_\p \leq 2 \aand r_\f=\frac{1}{2} ~\longrightarrow~ r_\p=\frac{3}{2}~.
\ea
The dimensions (UV and IR) of the Faddeev-Popov ghosts ($c$ and $\xi$) and antighosts ($\bc$ and $\bx$) are fixed, by considering 
the propagotors (\ref{propkcbc}), such that:
\ba
&& d_c + d_{\bc} \geq 1 \aand d_\xi + d_{\bx} \geq 1~; \\
&& r_c + r_{\bc} \leq 1 \aand r_\xi + r_{\bx} \leq 1~.
\ea 
Also, assuming that the BRS operator $s$ (\ref{BRS}) is dimensionless and bearing in mind 
that the coupling constant $g$ has dimension $(\rm{mass})^{\frac{1}{2}}$, we get the following results for the ghosts, antighosts and
$\r$ field:
\ba
&& d_c=-\frac{1}{2} ~,~~ d_{\bc}=\frac{3}{2} ~,~~ d_\xi=-\frac{1}{2} ~,~~ d_{\bx}=\frac{3}{2} \aand d_\r=-\frac{1}{2}~; \\
&& r_c=-\frac{1}{2} ~,~~ r_{\bc}=\frac{3}{2} ~,~~ r_\xi=-\frac{1}{2} ~,~~ r_{\bx}=\frac{3}{2} \aand r_\r=-\frac{1}{2}~.
\ea 
Finally, through the action of the antifields (\ref{sigmaext}), and the UV and IR dimensions of the fields fixed previously, it 
follows that
\ba
&& d_{A^*}=\frac{5}{2} ~,~~ d_{\f^*}=\frac{5}{2} ~,~~ d_{\r^*}=\frac{7}{2} ~,~~ d_{c^*}=\frac{7}{2} \aand d_{\xi^*}=\frac{7}{2}~; \\
&& r_{A^*}=\frac{5}{2} ~,~~ r_{\f^*}=\frac{5}{2} ~,~~ r_{\r^*}=\frac{7}{2} ~,~~ r_{c^*}=\frac{7}{2} \aand r_{\xi^*}=\frac{7}{2}~.
\ea

In summary, the UV and IR dimensions, $d$ and $r$ respectively, the ghost numbers, $\F\Pi$, of all fields are collected in Table~\ref{dimensions}.

\section{Slavnov-Taylor identity, ghost and antighost equations and Ward identities} 

This subsection is devoted to establish the Slavnov-Taylor identity, ghost and antighost equations, and two 
hidden rigid symmetries.
The BRS invariance of the action $\G^{(0)}$ (\ref{totalaction}) is expressed through the Slavnov-Taylor 
identity
\ba
\cs(\G^{(0)})&=&{\mbox{Tr}}\int d^3x~\Bigg\{ \frac{\d\G^{(0)}}{\d A^*_\m} \frac{\d\G^{(0)}}{\d A^\m}    
+ \frac{\d\G^{(0)}}{\d \f^*_\m} \frac{\d\G^{(0)}}{\d \f^\m} 
+ \frac{\d\G^{(0)}}{\d \r^*} \frac{\d\G^{(0)}}{\d \r}
+ \frac{\d\G^{(0)}}{\d c^*} \frac{\d\G^{(0)}}{\d c}
+ \frac{\d\G^{(0)}}{\d \xi^*} \frac{\d\G^{(0)}}{\d \xi} \nonumber\\
&+& b \frac{\d\G^{(0)}}{\d \bc} + \p \frac{\d\G^{(0)}}{\d \bx}\Bigg\}=0~,\label{Slavnov}
\ea
which translates, in a functional way, the invariance of the classical model under the BRS symmetry. It is 
suitable to define, for later use, the linearized Slavnov-Taylor ($\cs_{\G^{(0)}}$) operator as below
\ba
\cs_{\G^{(0)}}&=&{\mbox{Tr}}\int d^3x~\Bigg\{ \frac{\d\G^{(0)}}{\d A^*_\m} \frac{\d}{\d A^\m}    
+ \frac{\d\G^{(0)}}{\d A^\m} \frac{\d}{\d A^*_\m} 
+ \frac{\d\G^{(0)}}{\d \f^*_\m} \frac{\d}{\d \f^\m} 
+ \frac{\d\G^{(0)}}{\d \f^\m} \frac{\d}{\d \f^*_\m}
+ \frac{\d\G^{(0)}}{\d \r^*} \frac{\d}{\d \r}
+ \frac{\d\G^{(0)}}{\d \r} \frac{\d}{\d \r^*} \nonumber\\
&+& \frac{\d\G^{(0)}}{\d c^*} \frac{\d}{\d c}
+ \frac{\d\G^{(0)}}{\d c} \frac{\d}{\d c^*}
+ \frac{\d\G^{(0)}}{\d \xi^*} \frac{\d}{\d \xi} 
+ \frac{\d\G^{(0)}}{\d \xi} \frac{\d}{\d \xi^*}
+ b \frac{\d}{\d \bc} + \p \frac{\d}{\d \bx}\Bigg\}~.
\ea
Another identities, the ghost equations,
\ba
\cg_{\rm I}\G^{(0)}\equiv\frac{\d\G^{(0)}}{\d \bc}
+\pa^\m\frac{\d\G^{(0)}}{\d A^{*\m}}=0~,\\
\cg_{\rm II}\G^{(0)}\equiv\frac{\d\G^{(0)}}{\d \bx}
+\pa^\m\frac{\d\G^{(0)}}{\d \f^{*\m}}=0~,
\ea
follow from the gauge-fixing conditions, 
\ba
\frac{\d \G^{(0)}}{\d b}= \pa^\m A_\m + \a b~,\label{gaugefixing1a}\\
\frac{\d \G^{(0)}}{\d \p}= \pa^\m \f_\m + \b \p~,\label{gaugefixing2a}
\ea
and the Slavnov-Taylor identity (\ref{Slavnov}), 
meaning that $\G^{(0)}$ depends on the antighosts, $\bc$ and $\bx$, and the antifields, $A^{*\m}$ and $\f^{*\m}$, 
through the combinations 
\be
{\wt A}^*_\m=A^*_\m+\pa_\m\bc \aand 
{\wt \f}^*_\m=\f^*_\m+\pa_\m\bx~.
\ee

The Jackiw-Pi model presents two antighost equations, they are listed as below:
\ba
\ov\cg_{\rm I}\G^{(0)}&\equiv&\int d^3x~\biggl\{\frac{\d\G^{(0)}}{\d c}
-g\biggl[\bc,\frac{\d\G^{(0)}}{\d b}\biggr]-g\biggl[\bx,\frac{\d\G^{(0)}}{\d \p}\biggr]\biggr\}={\ov\D}_{\rm I}~, \label{antighost1} \\
{\mbox{where}}~~{\ov\D}_{\rm I}&\equiv&-g\int d^3x~\bigl\{[A^*_\m,A^\m]+[\f^*_\m,\f^\m]+[\r^*,\r]-[c^*,c]-[\xi^*,\xi]
+ \a[\bc,b]+\b[\bx,\p]\bigr\}~;\label{delta1}\\
\ov\cg_{\rm II}\G^{(0)}&\equiv&\int d^3x~\biggl\{\frac{\d\G^{(0)}}{\d \xi}
-g\biggl[\bx,\frac{\d\G^{(0)}}{\d b}\biggr]\biggr\}={\ov\D}_{\rm II}~,\label{antighost2}\\
{\mbox{where}}~~{\ov\D}_{\rm II}&\equiv&-g\int d^3x~\bigl\{[\f^*_\m,A^\m]-[\xi^*,c]-\frac{\r^*}{g}
+ \a[\bx,b]\bigr\}~.\label{delta2}
\ea
It should be noticed, for the sake of further quantization~\cite{jackiwpiquantum}, that the breakings, ${\ov\D}_{\rm I}$ and ${\ov\D}_{\rm II}$, being nonlinear in the quantum fields will be subjected to renormalization. An interesting issue in Yang-Mills theories is that the Landau gauge~\cite{landaugauge} has very special features as compared to a generic linear gauge. This is due to the existence, besides the Slavnov-Taylor identity, of 
another identity, the antighost equation~\cite{blasi-piguet-sorella}, which controls the dependence of the theory on the ghost $c$. In particular, 
this equation implies that the ghost field $c$ and the composite $c$-field cocycles in the 
descent equations have vanishing anomalous dimension, allowing the algebraic proof~\cite{piguet} of the Adler-Bardeen 
nonrenormalization theorem~\cite{adler-bardeen} for the gauge anomaly. Back to the Jackiw-Pi model we are considering here, in the case of the general linear covariant gauges, (\ref{gaugefixing1}) and (\ref{gaugefixing2}), the right-hand sides of the equations, (\ref{antighost1}) and 
(\ref{antighost2}), are nonlinear in the quantum fields due to the presence of the terms, 
$\int d^3x~\a[\bc,b]$ and $\int d^3x~\b[\bx,\p]$, and $\int d^3x~\a[\bx,b]$, respectively. Therefore, 
the breakings, ${\ov\D}_{\rm I}$ (\ref{delta1}) and ${\ov\D}_{\rm II}$ (\ref{delta2}) have to be renormalized, 
which could spoil the usefulness of the antighost equations, by this reason, bearing in mind later renormalization of the model, we adopt from 
now on the Landau gauge $\a=\b=0$. 

As another feature of the Landau gauge, the following Ward identities for the rigid symmetries stem from the Slavnov-Taylor identity 
(\ref{Slavnov}) and the antighost equations (\ref{antighost1}) and (\ref{antighost2}) with $\a=\b=0$:
\ba
\cw_{\rm I}^{\rm rig}\G^{(0)}&=&0~,~~
{\mbox{where}} \label{rigid1} \nonumber\\
\cw_{\rm I}^{\rm rig}&\equiv& -g \int d^3x~\biggl\{ \biggl[A^\m,\frac{\d}{\d A^\m}\biggr] + 
\biggl[\f^\m,\frac{\d}{\d \f^\m}\biggr] + \biggl[\r,\frac{\d}{\d \r}\biggr] + 
\biggl[b,\frac{\d}{\d b}\biggr] + \biggl[\p,\frac{\d}{\d \p}\biggr] + \biggl[c,\frac{\d}{\d c}\biggr] + 
\biggl[\xi,\frac{\d}{\d \xi}\biggr] + \nonumber\\
&+& \biggl[\bc,\frac{\d}{\d \bc}\biggr] + \biggl[\bx,\frac{\d}{\d \bx}\biggr] + 
\biggl[A^*_\m,\frac{\d}{\d A^*_\m}\biggr] + \biggl[\f^*_\m,\frac{\d}{\d \f^*_\m}\biggr] + 
\biggl[\r^*,\frac{\d}{\d \r^*}\biggr] + \biggl[c^*,\frac{\d}{\d c^*}\biggr] + \biggl[\xi^*,\frac{\d}{\d \xi^*}\biggr]
\biggr\}~;\\
\cw_{\rm II}^{\rm rig}\G^{(0)}&=&0~,~~
{\mbox{where}} \label{rigid2} \nonumber\\
\cw_{\rm II}^{\rm rig}&\equiv& -g\int d^3x~\biggl\{ \biggl[A^\m,\frac{\d}{\d \f^\m}\biggr] + 
\biggl[\p,\frac{\d}{\d b}\biggr] + \biggl[c,\frac{\d}{\d \xi}\biggr] + \biggl[\bx,\frac{\d}{\d \bc}\biggr] + 
\biggl[\f^*_\m,\frac{\d}{\d A^*_\m}\biggr] + \biggl[\xi^*,\frac{\d}{\d c^*}\biggr] + \frac{1}{g}\frac{\d}{\d \r} \biggr\}~.
\ea

\subsection{Operatorial algebra} 

All operators introduced previously satisfy the following off-shell 
algebra for any functional $\ck$ with even Faddeev-Popov charge:
\begin{enumerate}
\item Slavnov-Taylor operator identities:
\ba
&&S_{\ck}S(\ck)=0~\forall~\ck~,~~S_{\ck}S_{\ck}=0~{\rm if}~S(\ck)=0~,\nonumber\\
&&\frac{\d S(\ck)}{\d b}-S_{\ck}\left(\frac{\d\ck}{\d b}-\pa^\m A_\m\right)=
\cg_{\rm I}(\ck)~,~~
\frac{\d S(\ck)}{\d\p}-S_{\ck}\left(\frac{\d\ck}{\d \p}-
\pa^\m \f_\m\right)=\cg_{\rm II}(\ck)~,\nonumber\\
&&\cg_{\rm I} S(\ck)+S_{\ck}\cg_{\rm I}(\ck)=0~,~~
\cg_{\rm II} S(\ck)+S_{\ck}\cg_{\rm II}(\ck)=0~,\nonumber\\
&&\ov\cg_{\rm I} S(\ck)+S_{\ck}(\ov\cg_{\rm I}(\ck)-{\ov\D}_{\rm I})=\cw_{\rm I}(\ck)~,~~
\cw_{\rm I} S(\ck)-S_{\ck}\cw_{\rm I}(\ck)=0~,\nonumber\\
&&\ov\cg_{\rm II} S(\ck)+S_{\ck}(\ov\cg_{\rm II}(\ck)-{\ov\D}_{\rm II})=\cw_{\rm II}(\ck)~,~~
\cw_{\rm II}S(\ck)-S_{\ck}\cw_{\rm II}(\ck)=0~;
\label{algebra1}
\ea
\item Other identities:
\ba
&&\ov\cg_{\rm I}^a(\ov\cg_{\rm I}^b(\ck)-{\ov\D}_{\rm I}^b)+\ov\cg_{\rm I}^b(\ov\cg_{\rm I}^a(\ck)
-{\ov\D}_{\rm I}^a)=0~,~~
\cw_{\rm I}^a\cw_{\rm I}^b(\ck)-\cw_{\rm I}^b\cw_{\rm I}^a(\ck)=0~,\nonumber\\
&&\ov\cg_{\rm I}^a\cw_{\rm I}^b(\ck)-\cw_{\rm I}^a(\ov\cg_{\rm I}^b(\ck)-{\ov\D}_{\rm I}^b)=0~,\nonumber\\
&&\ov\cg_{\rm I}^a(\ov\cg_{\rm II}^b(\ck)-{\ov\D}_{\rm II}^b)+\ov\cg_{\rm II}^a(\ov\cg_{\rm I}^b(\ck)-{\ov\D}_{\rm I}^b)=0~,~~
\ov\cg_{\rm I}^a\cw_{\rm II}^b-\cw_{\rm II}^a(\ov\cg_{\rm I}^b(\ck)-{\ov\D}_{\rm I}^b)=0~,\nonumber\\
&&\ov\cg_{\rm II}^a\cw_{\rm I}^b(\ck)-\cw_{\rm I}^a(\ov\cg_{\rm II}^b(\ck)-{\ov\D}_{\rm II}^b)=0~,~~
\cw_{\rm I}^a\cw_{\rm II}^b(\ck)-\cw_{\rm II}^a\cw_{\rm I}^b(\ck)=0~,\nonumber\\
&&\ov\cg_{\rm II}^a(\ov\cg_{\rm II}^b(\ck)-{\ov\D}_{\rm II}^b)+\ov\cg_{\rm II}^b(\ov\cg_{\rm II}^a(\ck)
-{\ov\D}_{\rm II}^a)=0~,~~
\cw_{\rm II}^a\cw_{\rm II}^b(\ck)-\cw_{\rm II}^b\cw_{\rm II}^a(\ck)=0~,\nonumber\\
&&\ov\cg_{\rm II}^a\cw_{\rm II}^b(\ck)-\cw_{\rm II}^a(\ov\cg_{\rm II}^b(\ck)-{\ov\D}_{\rm II}^b)=0~,\nonumber\\
&&[\cp_\m,\Q]=0~\forall\Q\in\{S_{\ck},\cg_{\rm I},\cg_{\rm II},\ov\cg_{\rm I},\ov\cg_{\rm II},\cw_{\rm I},\cw_{\rm II},\cp_\m\}~,
\label{algebra2}
\ea
\end{enumerate}
where $\cp_\m$ is the Ward operator associated to translations:
\be
\cp_\m=\sum_\vf{\mbox{Tr}}\int d^3x~\pa_\m\vf\frac{\d}{\d \vf}~,
\ee
and $\vf$ are all the fields contained in the action (\ref{totalaction}).
The first group of identities involving the Slavnov-Taylor operator 
given by (\ref{algebra1}) are those which yield the conditions (the 
well-known Wess-Zumino consistency condition is one of them) to be 
satisfied by the quantum breaking of the Slavnov-Taylor identity 
(\ref{Slavnov}) allowed by the Quantum Action Principle \cite{piguet}.

\begin{table}[t]
\begin{center}
\begin{tabular}{|c||c|c|c|c|c|c|c|c|c|c|c|c|c|c|c|c|}
\hline
& $A_\m$ & $\f_\m$ & $\r$ & $b$ & $\p$ & $c$ & $\xi$ & ${\bc}$ & ${\bx}$ 
& $A^{*\m}$ & $\f^{*\m}$ & $\r^*$ & $c^*$ & $\xi^*$ & $g$ & $m$\\ 
\hline\hline
$d$ & $1/2$ & $1/2$ & $-1/2$ & $3/2$ & $3/2$ & $-1/2$ & $-1/2$ & $3/2$ & $3/2$ & $5/2$ & $5/2$ & $7/2$ & $7/2$ & $7/2$ & $1/2$ & $1$\\ \hline
$r$ & $1/2$ & $1/2$ & $-1/2$ & $3/2$ & $3/2$ & $-1/2$ & $-1/2$ & $3/2$ & $3/2$ & $5/2$ & $5/2$ & $7/2$ & $7/2$ & $7/2$ & $1/2$ & $1$\\ \hline
$\Phi\Pi$ & $0$ & $0$ & $0$ & $0$ & $0$ & $1$ & $1$ & $-1$ & $-1$ & $-1$ & $-1$ & $-1$ & $-2$ & $-2$ & $0$ & $0$\\ \hline
\end{tabular}
\end{center}
\caption[]{Ultraviolet dimension ($d$), infrared dimension ($r$) and ghost number ($\Phi \Pi$)\label{dimensions}.}
\end{table}

\section{Conclusions}

The Jackiw-Pi model which generates a mass gap preserving parity in three space-time dimensions was presented here. 
The BRS symmetry of the model was established and all the difficulties found out in the literature concerning the gauge-fixing 
were by-passed. At the tree-level the propagators were computed, the spectrum consistency (causality and unitarity) has been 
verified and we conclude that the Jackiw-Pi model are free from tachyons and ghosts. By the asymptotical behaviour of the propagators together 
with the BRS transformations, the ultraviolet and infrared dimensions of all the fields were fixed. Also, BRS invariance and Slavnov-Taylor identity together with the antighost equations, in the Landau gauge, allowed to find out two rigid symmetries, moreover, the operatorial algebra which defines the model has been presented. An important issue to notice is 
that, as we have shown, the Jackiw-Pi even-parity mass term is BRS invariant up to a total derivative, {\it i.e.}, 
it is not local BRS invariant. Therefore, it could be conjectured that, at the quantum level, the $\b$-function 
associated to the mass parameter $m$, $\b_m$, should be zero, $\b_m=0$~\cite{delcima-franco-helayel-piguet, barnich}. Moreover, this fact would indicate the perturbatively ultraviolet finiteness of the Jackiw-Pi model, which is now under investigation~\cite{jackiwpiquantum} 
in the framework of the algebraic renormalization scheme.

\subsection*{Acknowledgements}
In honor of the 70th birthday of Prof. Olivier Piguet. Also, the author dedicates this work to his kids, Vittoria and Enzo, 
and to his mother, Victoria. Thanks are also due to Daniel H.T. Franco for discussions and encouragement.

\end{document}